\documentclass[letterpaper, 10 pt, conference]{ieeeconf}  

\IEEEoverridecommandlockouts                              
\usepackage{hyperref}

\title{\LARGE \bf
A Critical Survey of Privacy Infrastructures
}

\author{B H Priyanka$^{1}$ and Ravi Prakash$^{2}$
\thanks{$^{1}$College of Computer and Information Science, Northeastern 	         	         University, Boston, MA, USA,
              email: bhadravathihalesh.p@husky.neu.edu}%
\thanks{$^{2}$ Bengaluru, Karnataka, INDIA,
               email: ravi\_1010@hotmail.com}%
}

\begin{document}

\maketitle
\thispagestyle{empty}
\pagestyle{empty}

\begin{abstract}
Over the last two decades, the scale and complexity of the Internet and its associated technologies built on the World Wide Web has grown exponentially with access to Internet as a facility occupying a prime place with other amenities of modern lives. In years to come, usage of Internet may unravel more pleasant surprises for us as far as novelty in its usage is concerned. As a democratic function of Internet, and relying on the open model on which it has been built, there has been concerted efforts in the direction of privacy protection and use of privacy enhancing tools which have gained tangible traction. Innovation in use of VPN, TLS/SSL and cryptographic tools are a testimony to it. Another popular tool is \textit{Tor} which has gained widespread popularity as it is being increasingly used by anonymity seeking users to effectively maintain their discretion while surfing the web. \\

However, there is a darker side to increased proliferation of Internet in our everyday routine. We are certainly not living in a utopian age and there are potentials of misuse of Internet as well. Across every nook and cranny of Internet's sprawling virtual world, there are cyber criminals lurking n dangerous alleys to use the very same Internet as malevolent tool to abuse it and cause financial, physical and social harm to ordinary people. Failing to manage the widespread spawning of World Wide Web has rendered it weak against misuse. In last few decades especially, Internet has been inundated with malware, ransomware, viruses, Trojans, illegal spy tools and what not created with malignant sentiments. In this paper, we will analyze few of the subverting privacy infrastructures. \\
\end{abstract}

\section{INTRODUCTION}
A cursory look at the major security breaches in the last few years like hacking of big multinational firms (Sony, Amazon etc.), online payment agencies, stealing of credit card information have revealed a glaring truth about vulnerability of even most secure digital infrastructure to be compromised and subsequently cause massive financial loss in terms of millions of dollars. Modus operandi of cyber criminals is increasingly becoming stealthy and pervasive. As a natural evolution of internet, there has been progress in privacy enhancing technologies like Tor, VPN, etc. which have potential to be used in maintaining genuine anonymity, but at the same time are facing threats of subversion. In this paper, we analyze following topics and their perceived threats:\\

1. Botnets\\

2. Tor\\

3. VPN\\

4. Wi-Fi\\

While discussing each of the above topics, we present a brief background, current scenario and future assessment. Also, in the conclusion part, we suggest our estimates with respect to mitigating their threats.\\

\section{BOTNETS}

\subsection{Background}

A \textit{Botnet} is a generic name given to a cluster of computers which is managed by an adversary for mala fide usage. So, essentially a botnet is a network of compromised computers which are called \textit{bots}. Internet Relay Channels (IRC) which rely on text based communication amongst a group of peers were in vogue during the dawn of Internet era. It was only natural that they were the debut targets of botnet cyber criminals.\\

In 1999, bots were created to establish a connection to the IRC channel, which in turn was capable of triggering an IRC client communication towards IRC server. Subsequently, bots could control the channel itself and hence spy on the entire network. The bot, would then connect to a certain chat room which could  then entice multiple bots to join the channel, thereby enabling the botmaster (one who controls the actions of several bots in the network) to cause threats to the IRC servers. This behavior is still evident in the W32/Vulcanbot attack on websites of the human-rights activists.\\

Botnets shone into limelight, when a series of distributed Denial of Service attacks were launched on popular websites like Yahoo, Ebay, Dell and Amazon in the year 2000. Eventually, researchers and the security experts denied the fact that these DoS attacks were actually caused by a botnet, but deduced the fact that the botnets pose a serious threat to the soundness of the Internet in the future. Few of the renowned bots are SDBot, SpyBot, Conficker, Zeus, Waledac, Mariposa and Kelihos.\\

\subsection{Threats posed by Botnets}

Botnets have been key players in subverting the privacy infrastructure and the cryptography mechanisms used by Internet users to ensure privacy. Major threats being Distributed DoS which bombards the server with multiple requests eventually leading to the exhaustion of the bandwidth and the resources, sending out enormous amount of spams to the users, accessing the systems comprising of critical data to fetch the private/secret information of the users and modifying the source code to inject additional malware to exploit the vulnerabilities exposed.\\

\subsection{Bunito Botnet}

Recent works on analysis of \textit{Bunito Botnet} show that cyber criminals had employed a server to widely distribute the botnet infected proxies to the users and have utilized this as a main source of capital. There has been a considerable increase in the percentage of users who have been infected with this Bunito botnet in the last decade. Users primarily use the VPN service to connect to a private network securely through a VPN service provider and are oblivious of the fact that the VPN service providers through which they connect are fraudulent and are under the nefarious influence of Bunito botnet. VIP72 is a well-known VPN service that has gained disrepute due to the kind of anonymity they provide with respect to the proxies and also because of their close knit connection with the Bunito botnet.\\

\subsubsection{Botnets \& TLS}

TLS/SSL are usually used by legitimate users for genuine security needs. Security is achieved by employing cryptographic functions at transport layer to establish secure communication channel. However, creators of botnets have been no less enterprising in embracing new security apparatus like TLS/SSL for abuse. A botnet named \textit{Rustock} was unearthed using same TLS encryption mechanism for spreading huge amount of spams. Security firms across the world have observed gradual increase in TLS based spam.\\

\subsubsection{Mobile Botnets}

There has been unprecedented increase in users of smartphones, tablets and other smart digital devices which run on a plethora of Operating Systems like Android, Tizen, iOS and Windows. Hence, it is not amusing to see mobile platforms as next logical target of botnets. Mobile botnets work exactly as the botnets which infect the computers, mainly exploiting the vulnerabilities found in the mobile Operating System. The hacker uses the privileges of the root user to send text messages, calls, media, etc. RootSmart, BBproxy, Android snooping, DroidDream are some of the several well-known mobile botnets which have gained attention in the recent years.

\section{TOR}

\textit{Tor}, an acronym for \textit{The Onion Router}, is a freely available software which provides enhanced privacy features to the users through a network of computers which persevere hard to reduce or completely eliminate digital footprint of the users and vanquishes traces of the communication that the user makes, be it with another computer or just browsing on the Internet. Tor, unlike botnet did not come into existence because of any criminal pursuit, but due to genuine academic interest.\\

Its public history dates back to the year 1995, when the US Naval Research Lab started analyzing and funding \textit{Onion Routing} which gradually led to open sourcing of the \textit{Tor} project and was made available to general public. Subsequently, widespread reception of \textit{Tor} software took place by anonymity seeking users who were actively looking for a viable and reliable alternative.\\

It was the democratic and secular ethos enshrined in the modelling, functioning and usage of \textit{Tor} which made it a favorite of freedom seeking activists who despised unsolicited snooping on their online activities. In March 2011, The Tor Project received the \textit{Free Software Foundation}'s 2010 Award for Projects of Social Benefit. The citation read, \textit{"Using free software, Tor has enabled roughly 36 million people around the world to experience freedom of access and expression on the Internet while keeping them in control of their privacy and anonymity."} Its network has proved pivotal in dissident movements in many countries.\\

Crypto anarchism as a movement had huge gains from using \textit{Tor} in areas like political dissidence, whistleblowing, cyber activism etc. It helped people associated with such movements to evade censorship, prosecution and escape repressive regimes. Intelligence agencies also used \textit{Tor} model to evade espionage from non-friendly countries.\\

\subsection{Threats posed by Tor}

Given the alacrity with which nefarious activities take place in cyberspace, it was no wonder that criminals saw it as another avenue for plying their illegal trade. Abusers of \textit{Tor} have posed a severe threat to data and network integrity by performing a range of disruptive activities like sending emails with malware infected content and launching Distributed DoS attacks to the servers . Within the comfortable veil of secrecy, the dark side of \textit{Tor} gets even more darker.\\

Service providers like \textit{Freedom hosting} which promoted child abuse, \textit{Silk Road}, \textit{Black Market Reloaded}, \textit{The Armory} and \textit{The General Store} which illegally dealt with selling and buying of drugs, arms and ammunitions are few of the hidden sites associated with Tor.\\

Silk Road, a popular online market place for trade of illegal drugs was launched in the year 2001, which provided complete anonymity of both the sellers and buyers. It also sold numerous fake documents such as driver's license, hacked Amazon accounts, fake passports, hacked Netflix account details and fake credit card statements. Silk Road used \textit{Bitcoin} as a mode of its payment by its users, which is an innovative distributed crypto currency system which provides no traces about the payments done online. There was a tremendous surge in the transactions carried on Silk Road which made it become very much conspicuous to the security organizations in US.\\

\subsubsection{Tor as a Malware}

Researchers recently exposed malware that infected point-of-sale terminals at several dozen retailers in the US and other countries and successfully captured customers' payment card data. \textit{Chewbacca}, as the crime ware was called, removes the memory bits and dumps into files. It then uses regular expressions and other programming techniques to extract data that was copied from credit and debit cards. Chewbacca also captured sensitive data using a generic key logger.\\

\subsubsection{Tor coupled with insecure protocols}

Another glaring issue with Tor is that there is a prevalence of insecure protocols such as POP3 which use plaintext for communication. It has successfully demonstrated that it is possible to see passwords by logging exit traffic on routers.\\

\subsubsection{Tor with mobile applications}

As in case of prevalence of botnets on mobile applications, Tor has found both use and misuse on mobile platforms. There are plenty of mobile applications which ensure that all traffic is routed through Tor networks guaranteeing privacy. Applications like \textit{Orbot} and \textit{Orweb} to name a few are readily available.\\

Consistency in the usage of same Operating System like Android or iOS which use Tor for a long period of time, gives chance to the sophisticated cyber criminals to exploit the vulnerabilities in these platforms. The malware which infects the operating system of the mobile devices can gather all personal data of the users, be it contacts, images, any other sensitive data stored in the memory cards, hack the text messages, track the location of the user using the GPS technology and also the internal specifications of the device itself.

\section{VIRTUAL PRIVATE NETWORKS}
\subsection{Background}

Virtual Private Network (VPN) is one of the well-known privacy infrastructure that has been trending in the past few years. It is a private network that uses Internet to provide a secured data transmission across a channel, thereby having access to all the resources of the remote network directly. VPN, as a form of Wide Area Network(WAN) provides various services like file sharing, audio and video conferencing, email or instant messaging, is popular due to its cost effectiveness and its coverage area be it intranet or internet. Mainly used by the IT employees to remotely access the corporate websites, VPN is a convenient and efficient means of connecting to the private network of companies. Not only restricted to corporate companies, VPN is now a sophisticated and advanced privacy technology which any user can use.\\

\subsection{Threats posed by VPN}

\subsubsection{Hola VPN}

There have been quite a few instances of misuse of commercial VPN services. For example, a commercial private VPN \textit{hola} misused the resources of its clients without any disclosure. It was only after exposure, that the malpractice of the operators of hola vpn came into limelight where they sold connections of their users to another \textit{Tor} like vpn called Luminati.\\

\subsubsection{Heartbleed Virus}
\textit{Heartbleed} virus took IT sector by storm when it exposed chinks in the armour of OpenSSL implementation. IT firms across the world scurried to patch the bug as OpenSSL is widely used for security. VPN implementation built on OpenSSL are extremely vulnerable and it will take years to evaluate, assess and remedy the impact of Heartbleed virus.\\

\section{WIRELESS FIDELITY}

A Wireless Local-area Network (LAN) uses radio waves to connect devices. Wi-Fi technology brought radical changes in the way internet was accessed. Convenience, mobility, productivity, cost effectiveness were some of its hallmark technology which led to its wide adoption. Although, satisfactory security infrastructure is in place for Wi-Fi, however even Wi-Fi is not completely reliable.\\

Recently, security researchers managed to demonstrate how the unauthenticated data packets in the 802.11 wireless LAN protocol can be used as a covert channel to control malware on an infected computer. 802.11 protocol relies on clients and access points exchanging informational data packets before they authenticate or associate with each other, and this traffic is not typically monitored by network security devices. Tom Neaves, a managing consultant at Trustwave, developed a proof-of-concept tool called Smuggler that leverages these packets, known as wireless management frames, to communicate with malware.\\

\section{CONCLUSIONS AND THE WAY FORWARD}

\subsection{\textbf{Botnet}}
Botnets will continue to dominate security landscape in years to come for sure. The need of the hour is to scrupulously plug the existing loopholes to decimate the threats of botnets. Cyber criminals will certainly innovate newer mechanism to spread the botnet but a vigilant security model in place which takes care of intrusion prevention and detection will alleviate the risks in long term.\\

\subsection{\textbf{Tor}}

Popularity of \textit{Tor} is ever increasing and its supporters are ardent advocates of the internet freedom. Hence, prognosis of Tor project looks good. But, the Tor infrastructure must be regularly updated to deal with new contingencies and impinge on its anonymity model. Much to the chagrin of repressive regimes who don't prefer activists using \textit{Tor} to voice their opinion, are the one looing for new way to scuttle or hack the Tor infrastructure. Also, off late, hackers have increasingly made efforts to infiltrate Tor and spread malware through it. As it is an open source project, Tor needs a vibrant and thriving community of volunteers which can keep it going strongly and faithfully.\\

\subsection{\textbf{VPN}}

VPN has become the de facto standard for secure access to digital infrastructure. There are wide range of VPN deployments across the world using a gamut of technologies. As expected with any popular technology, VPNs are also prone to vulnerabilities as Heartbleed virus which was mentioned above. There is a pressing need to create global standard for VPN which adheres to stringent security ideals in implementation and in practice.\\

\subsection{\textbf{Wi-Fi}}
Use of Wi-Fi and its complimentary technologies have spread like wildfire. Organizations across the world are exhorting their members to move to BYOD model. Hence, it becomes imperative to ensure that Wi-Fi access is safe and authorized.
Also, there is scope for newer techniques and models to be used in Wi-Fi for ensuring secure network alarms.\\

\addtolength{\textheight}{-12cm}   



\end{document}